\documentstyle[11pt,tbrown_nhst,twoside,epsf]{article}
\begin{document}
\title{100 Times Faster and 3 Times Sharper: Background-Dominated Observations
of Stellar Populations with an 8-meter Optical-UV Space Telescope}
 \author{Thomas M. Brown}
\affil{Space Telescope Science Institute, 3700 San Martin Drive, Baltimore,
MD, 21218}

\begin{abstract}
An 8 m successor to the Hubble Space Telescope (HST) would make
incredible gains in the study of stellar populations, especially in
the Local Group.  If diffraction-limited at 0.5 $\mu$m, the ``Next HST''
could produce high-resolution imaging of faint sources over a wide
field in 1 percent of the time needed with the HST.  With these
capabilities, photometry of the ancient main sequence could be
obtained for many sight-lines through Local Group galaxies, thus
determining directly the ages of their structures and providing a
formation history for the Local Group populations.
\end{abstract}

\section{Introduction}

One of the primary quests of observational astronomy is measuring the
formation history of giant galaxies. Recently, renewed interest in
formation via accretion of dwarf galaxies has been sparked by the
discoveries of the Sagittarius dwarf galaxy falling into the Milky Way (Ibata
et al. 1994) and of a tidal stream in the Andromeda halo (Ibata et
al. 2001). With large investments of Hubble Space Telescope (HST)
time, color-magnitude diagrams that reach the ancient main sequence
can be constructed for selected fields in galaxies of the Local Group,
thus providing accurate ages for their structures via the same
techniques traditionally used to date the populations in Galactic
globular clusters.  Unfortunately, these studies are limited by the
long integration times needed to reach the main sequence at the
distance of Andromeda (the nearest spiral to our own), and by the
stellar crowding that can be addressed at the HST resolution. However,
an 8 m optical-UV space telescope, diffraction-limited at 0.5
$\mu$m, would crack this field wide open, because of a simple concept
that is often overlooked in discussions of an HST successor: The time
to reach a given signal-to-noise for background-dominated photometry
of point sources scales as aperture to the fourth power, for a
telescope that is diffraction-limited at a given wavelength.

\section{Background-Dominated Photometry}

At a given wavelength, the width of a resolution element decreases
linearly with the aperture diameter ($D$), and thus the area of a
resolution element on the sky decreases as $D^2$. Because the collecting
area increases as $D^2$, the sky counts in a resolution element remain
constant as aperture increases. The signal-to-noise ratio ({\it S/N}) for
background-dominated photometry thus scales as: \\

$ {\it S/N} ~~ \propto ~~ 
{{S \times t \times ({D \over d})^2} \over 
{\sqrt{B \times t ~ + ~ S \times t \times ({D \over D_o} \times {d_o \over d})^2}}} ~~ \propto ~~
{{S \times t \times ({D \over d})^2} \over {\sqrt{B \times t}}} ~~ \propto ~~
S \times \sqrt{t} \times ({D \over d})^2
$ \\

\noindent
where the source rate is $S$, the sky background rate is $B$ ($B \gg
S$), the exposure time is $t$, and the source distance is $d$. Thus,
with other parameters fixed, the distance at which you can see a
source scales as $D$, the flux you can measure scales as $D^{-2}$,
and the exposure needed scales as $D^{-4}$. Current HST observing
programs studying faint stars would be 100 times faster with
an 8 m optical-UV space telescope, without assuming any advances
in optical coatings, detector efficiencies, etc. (see Figure 1).
These dramatic gains will be even larger than those realized by the
Next Generation Space Telescope (a 6 m telescope, diffraction
limited at 1 -- 2 $\mu$m, with higher background in its
near-IR channels). \\

\noindent
\parbox{3.5in}{\epsfxsize=3.5in \epsfbox{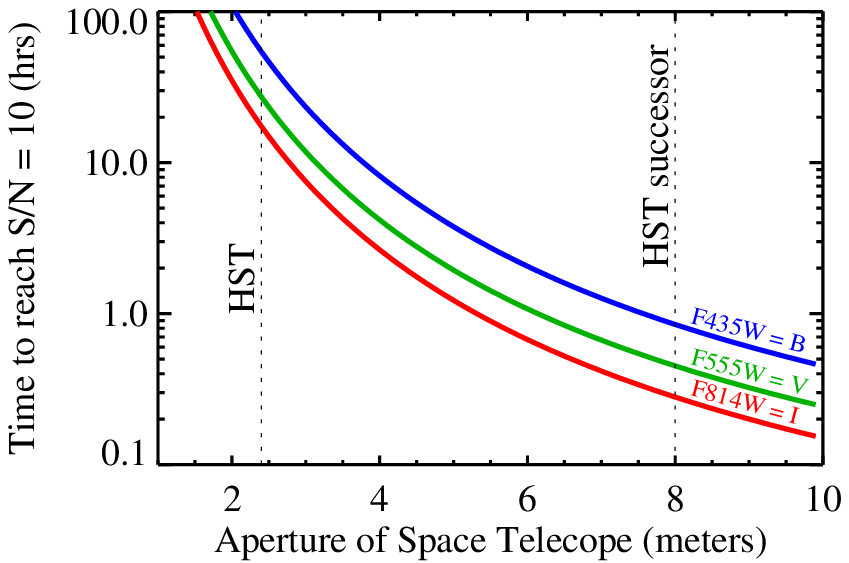}}

\vskip -2.1in \hskip 3.2in
\parbox{1.7in}{\small Figure 1. The exposure time needed to obtain photometry
of the Sun at the distance of Andromeda, assuming a telescope that is
diffraction-limited at 0.5 $\mu$m.  The Sun makes a good fiducial
because of its familiarity and its proximity to the MSTO in the
color-magnitude diagram of an old stellar population.}

\vskip 0.5in

Many difficult problems, requiring large investments of HST observing
time (more than 100 orbits), become trivial with an 8 m version of
the HST. The shorter integrations make it possible to study many
more sight lines through Local Group galaxies, and the superior
resolution allows investigation into structures that are much more
crowded. Because the exposure time for a given {\it S/N} is so much
faster, one can also study variability in regimes previously
unreachable. Examples of programs that would benefit greatly from a
larger optical-UV space telescope include age-dating Local Group
populations using the main sequence turnoff (MSTO), 
studying the white dwarf cooling curve in
Galactic globular clusters, and microlensing searches for dark matter.

\section{Simulations}

To dramatize the gains made in imaging of stellar populations, I have
created two simulations of imaging in the bulge of Andromeda, of which
only small subsections are shown in Figure 2. The first
simulation shows a combination of three bands from a
hypothetical HST imaging program using the Advance Camera for Surveys
(ACS) and its Wide Field Camera (WFC), assuming the native pixel size
(50 mas).  This false-color image combines F435W data (120 hrs) for the
blue channel, F606W data (42 hrs) for the green channel, and F814W
data (49 hrs) for the red channel. Although such deep exposures could
detect isolated stars on the ancient main sequence at this distance,
the crowding and surface brightness of the unresolved stars in the
bulge and disk would preclude detection of the MSTO.  The second
simulation shows the same field as imaged with the same bandpasses
using the ``Next HST'' (NHST; an 8 m version of the HST), assuming
1/10$^{\rm th}$ the exposure time; note that only 1/100$^{\rm th}$ of
the HST time is needed to match the HST {\it S/N}.  The NHST
simulation resolves the ancient MSTO. The NHST would likely have two
imaging cameras: a wide-field camera that undersamples the point
spread function (PSF) with an approximately 1 square degree field, and
a high-resolution camera that critically samples the PSF (7.5 mas
pixels) with a field-of-view comparable to the current ACS/WFC on HST
(204$^{\prime \prime} \times 204^{\prime \prime}$). The NHST
simulation uses the high-resolution camera, which provides the same
field, but with a PSF that is both critically sampled and 3.3
times narrower than the PSF in the HST simulation.

The simulations reproduce the surface brightness in a field about 8
arcmin from the center of Andromeda, where the bulge and disk
populations have approximately the same surface brightness (Baggett,
Baggett, \& Anderson 1998).  The full simulated images include a
globular cluster, young open cluster, and tidal stream from a
disrupted galaxy (10\% overdensity).  All of the components (disk,
bulge, stream, and clusters) were constructed using the isochrones of
Bertelli et al.\ (1994).  Figure 2 shows $7.1^{\prime \prime} \times
5.4^{\prime \prime}$ subsections in the vicinity of the Andromeda
globular cluster.

\section{Summary}

For background-dominated photometry of stellar populations, an 8 m
version of the HST would be 100 times faster than the current HST, and
provide 3 times the resolution.  Although the increased resolution
would allow photometry in fields that are currently impossible with
the HST, the reduction in exposure time is not an advance that should
be treated lightly.  Obviously, for a given field that can be resolved
with the HST, one can always argue that requesting one hundred orbits
of HST time would be easier than building an 8 m HST and observing
for one orbit.  However, a factor of 100 in exposure time allows one
to resolve the main sequence in dozens of sight-lines throughout the
Local Group, a program that is simply impossible with the HST, no
matter what fraction of its remaining lifetime is devoted to the
problem.  Such a program would allow the direct age determination of
structures (disks, bulges, halos) in Local Group galaxies, providing
a formation history for these galaxies and their components.

\noindent
\parbox{5.25in}{\epsfxsize=5.25in \epsfbox{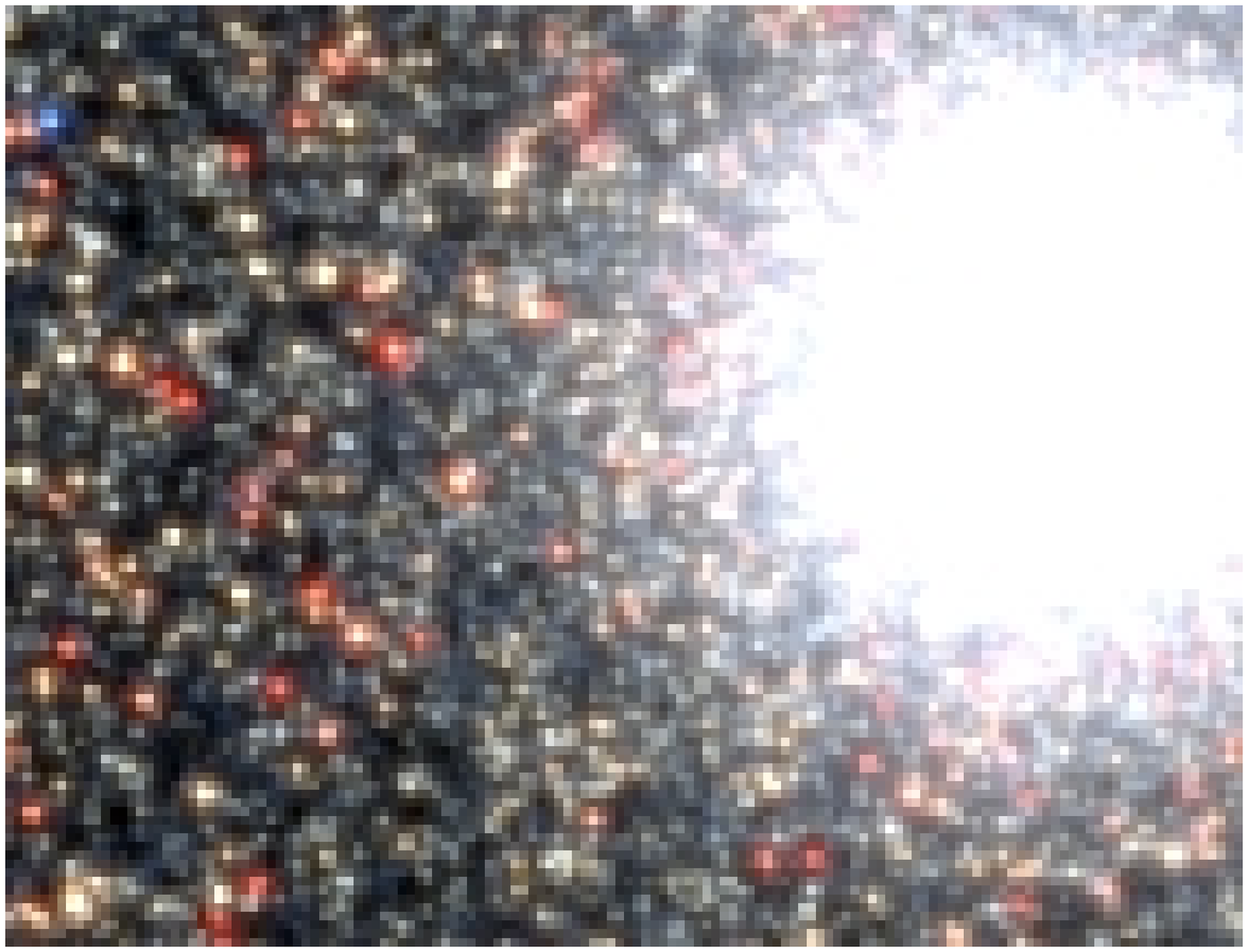}}

\noindent
\parbox{5.25in}{\epsfxsize=5.25in \epsfbox{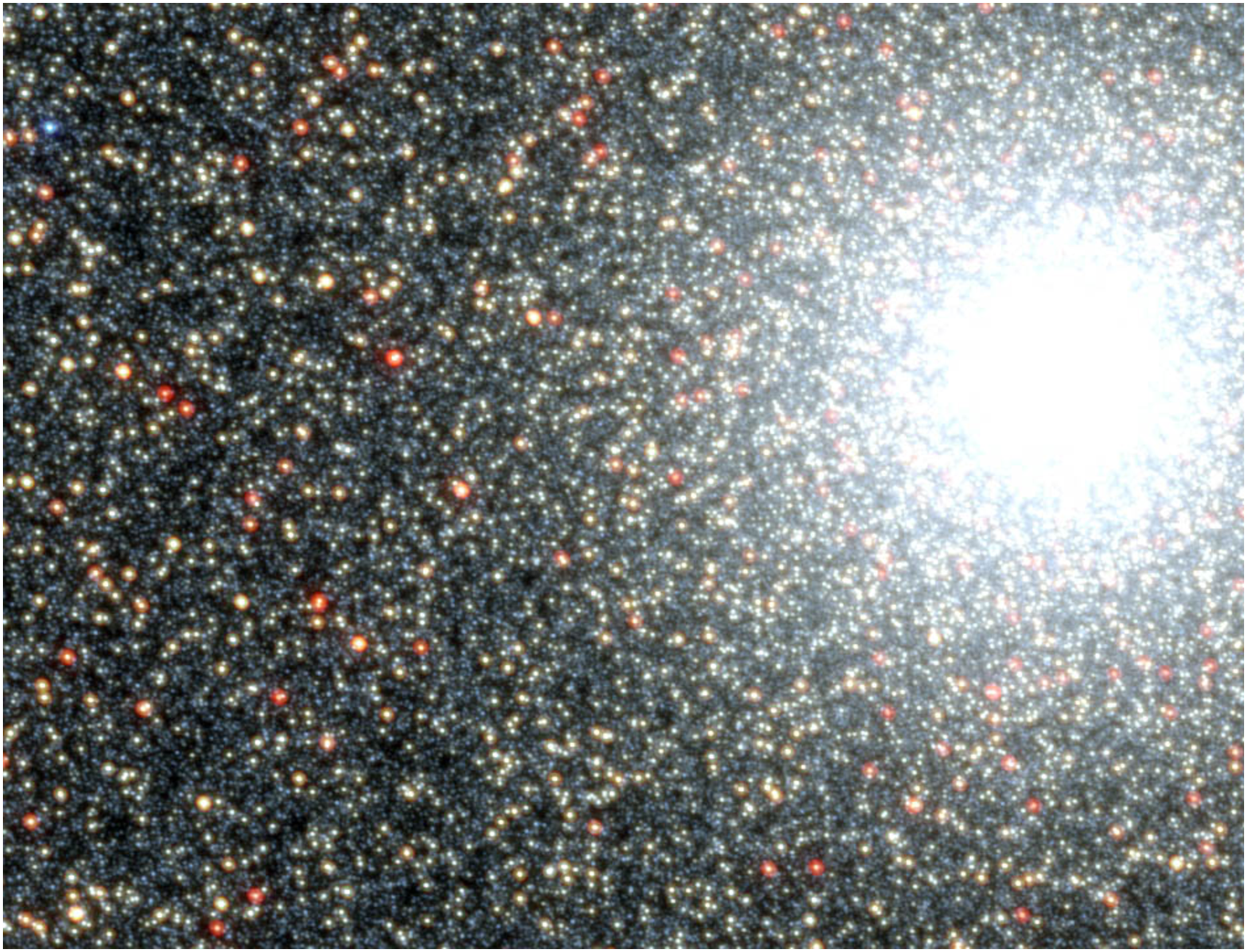}}

\noindent
\parbox{5.25in}{\small Figure 2. Simulated 
images in the bulge of Andromeda, near a globular cluster
(see text for details). {\it Top panel:} HST/ACS/WFC; {\it bottom panel:}
NHST.}

\end{document}